\documentclass [a4paper, 12pt]{article}
\usepackage{graphicx}
\usepackage[small]{subfigure,epsfig}

\usepackage {amsmath} \usepackage{amssymb} \usepackage{cite}

\begin{document}

\title{Analytical and numerical studying of the perturbed Korteweg--de Vries equation}
\author{Nikolay A. Kudryashov and Dmitry I. Sinelshchikov}
\date{Department of Applied Mathematics, National Research Nuclear University MEPhI (Moscow Engineering Physics Institute), 31 Kashirskoe Shosse, 115409 Moscow, Russian Federation}

\maketitle

\begin{abstract}
The perturbed Korteweg--de Vries equation is considered. This equation is used for the description of one--dimensional  viscous gas dynamics, nonlinear waves in a liquid with gas bubbles and nonlinear acoustic waves. The integrability of this equation is investigated using the Painlev\'{e} approach. The condition for parameters for the integrability of the perturbed Korteweg--de Vries equation equation is established. New classical and nonclassical symmetries admitted by this equation are found. All corresponding symmetry reductions are obtained. New exact solutions of these reductions are constructed. They are expressed via trigonometric and Airy functions. Stability of the exact solutions of the perturbed Korteweg--de Vries equation is investigated numerically.
\end{abstract}

\maketitle

\section{Introduction}
Nonlinear evolution equations play an important role in the modern mathematics and physics. There are several ``universal'' equations that are widely physical applicable and have remarkable mathematical properties (see e.g., Refs. \cite{Calogero1991,Leblond2008}). In this set of equations one can include the Korteweg--de Vries equation, the Burgers equation, the nonlinear Shrodinger equation. Such ``universal'' equations are very often obtained with the help of some asymptotic approach, for example, with the help of the reductive perturbation method (see e.g., Refs. \cite{Gardner1969,Taniuti1978,Leblond2008} ). If we consider the first order approximation in the asymptotic expansion we can obtain famous nonlinear evolution equations such as the Burgers and the Korteweg--de Vries equations. On the other hand, taking into account high order corrections in the reductive perturbation method one can obtain generalizations of the above mentioned equations. These equations can be considered as ``universal'' equations with high order corrections.

In this work we study the Burgers equation with high order corrections. This equation was obtained for the description of viscous gas dynamics and nonlinear acoustic waves in Refs. \cite{Fokas1996,Kraenkel1998}. In Refs. \cite{Kudryashov2013,Kudryashov2013a} it was shown that the Burgers equation with high order corrections can be used for the description of nonlinear waves in a liquid with gas bubbles. This equation has the form\cite{Kudryashov2013,Kudryashov2013a}:
\begin{equation}
\begin{gathered}
u_{t}+\alpha u u_{x}-\mu u_{xx}=\varepsilon\Bigg((2\mu\alpha_{2}+\mu\alpha+\nu)u u_{xx}+(2\mu\alpha_{1}+\mu\alpha+\nu)u_{x}^{2}-\\-\frac{\alpha(\alpha_{2}+2\alpha_{1})}{2}u^{2}u_{x}
-(\beta+\mu^{2}) u_{xxx}\Bigg),
\label{eq:ext_Burgers_1}
 \end{gathered}
\end{equation}
where $t$ is the non--dimensional time, $x$ is the non--dimensional Cartesian coordinate, $u$ is non--dimensional perturbation of gas--liquid mixture density, $\alpha$, $\beta$, $\nu$, $\mu$ are non--dimensional physical parameters, $\alpha_{1}$, $\alpha_{2}$ are arbitrary parameters introduced by the near--identity transformations (see. Ref. \cite{Kudryashov2013a}). Note that in fact Eq. \eqref{eq:ext_Burgers_1} is a two--parametric family of equations parameterized by $\alpha_{1}$ and $\alpha_{2}$.

It is known that at the derivation of the Burgers equation with high order corrections appears an obstacle to the integrability (see Refs. \cite{Fokas1996,Kraenkel1998,Zarmi2005,Kudryashov2013,Kudryashov2013a}). Eq. \eqref{eq:ext_Burgers_1} is integrable under a certain condition on the physical parameters in contrast to the Burgers equation. In the integrable case Eq. \eqref{eq:ext_Burgers_1} coincides with the Sarmo--Tasso--Olver equation which was investigated (see Refs. \cite{Olver1977,Tasso1996,Fokas2008,Kudryashov2009,Kudryashov2012}). However, the general case of the Burgers equation with high order corrections has not been studied previously. Thus it is an interesting problem to perform an analytical and numerical study of the Burgers equation with high order corrections without imposing any conditions on the physical parameters.

Below we show that Eq. \eqref{eq:ext_Burgers_1} can be transformed to a canonical form with a single parameter. This parameter can be considered as the bifurcation parameter, because there is only one value of this parameter corresponding to the integrable case of Eq. \eqref{eq:ext_Burgers_1}. The aim of this work is to analytically and numerically study the canonical form of Eq. \eqref{eq:ext_Burgers_1} at all values of the bifurcation parameter.

Let us remark that the canonical form of Eq. \eqref{eq:ext_Burgers_1} can be considered as a perturbed Korteweg--de Vries equation since
some of the leading terms of these equations are the same. Further we will refer to the canonical form of Eq. \eqref{eq:ext_Burgers_1} as a perturbed Korteweg--de Vries equation.

In order to investigate the canonical form of Eq. \eqref{eq:ext_Burgers_1} we apply the Painlev\'{e} and symmetry approaches. Using the Painlev\'{e} approach we illustrate that this equation is integrable only at a certain value of the bifurcation parameter. With the help of the symmetry approach we find both classical and nonclassical symmetries admitted by the perturbed Korteweg--de Vries equation. We construct corresponding symmetry reductions of this equation and obtain their exact solutions. We also numerically study the stability and dynamics of certain solutions of the perturbed Korteweg--de Vries equation. To the best of our knowledge our results are knew.

The rest of this work is organized as follows. In Sec. \ref{sec:2} we transform Eq. \eqref{eq:ext_Burgers_1} to the canonical form. The Painleve analysis of this equation is carried out as well. Symmetries of the perturbed Korteweg--de Vries equation are studied in Sec. \ref{sec:3}. We investigate both classical and nonclassical symmetries of this equation. Symmetry reductions of the perturbed Korteweg--de Vries equation are obtained in Sec. \ref{sec:4}. We construct exact solutions for the symmetry reductions of the perturbed Korteweg--de Vries equation in Sec. \ref{sec:4} as well. Sec. \ref{sec:5} is devoted to the numerical investigation of the perturbed Korteweg--de Vries equation. We give final remarks and conclusions in Sec. \ref{sec:6}.

\section{\label{sec:2}Transformation to the canonical form and Painleve analysis}

Let us show that Eq. \eqref{eq:ext_Burgers_1} can be simplified with the help of the scaling transformations and using arbitrariness of $\alpha_{1}$, $\alpha_{2}$. Indeed, assuming that $\alpha_{1}$ and $\alpha_{2}$ have the form
\begin{equation}
\alpha_{1}=\alpha_{2}=\frac{\mu\alpha+\nu}{\mu},
\label{eq:alpha_values}
\end{equation}
and substituting transformations
\begin{equation}
x^{'}=x+\frac{\alpha\mu}{6\varepsilon(\mu\alpha+\nu)}t, \quad
t^{'}=(\beta+\mu^{2})\varepsilon\,t,\quad
u=\frac{\beta+\mu^{2}}{\mu\alpha+\nu}\,u^{'}-\frac{\mu}{3\varepsilon(\mu\alpha+\nu)},
\label{eq:ext_B_transformation}
\end{equation}
into Eq. \eqref{eq:ext_Burgers_1} we have (primes are omitted)
\begin{equation}
u_{t}-3(u u_{x})_{x}+3\bar{\alpha}\,u^{2}u_{x}+u_{xxx}=0,
\label{eq:ext_Burgers_3}
\end{equation}
where
\begin{equation}
\bar{\alpha}=\frac{\alpha(\beta+\mu^{2})}{2\mu(\mu\alpha+\nu)}.
\label{eq:bar_alpha}
\end{equation}
Further we study Eq. \eqref{eq:ext_Burgers_3}.  Note that it seems appropriate to refer to Eq. \eqref{eq:ext_Burgers_3} as a perturbed Korteweg--de Vries equation as far as some of the leading terms of these equations are the same.

Let us investigate whether Eq. \eqref{eq:ext_Burgers_3} possesses the Painelve property for partial differential equations. In accordance with the algorithm by Weiss, Tabor and Carnevale\cite{Weiss1983} we present a solution of  Eq. \eqref{eq:ext_Burgers_3} in the form
\begin{equation}
u(x,t)=\phi^{p}\sum\limits_{j=0}^{\infty}u_{j}\phi^{j}, \quad u_{j}\equiv u_{j}(x,t),\quad \phi\equiv\phi(x,t).
\label{eq:WTC_expansion}
\end{equation}
The necessary condition for Eq. \eqref{eq:ext_Burgers_3} to possess the Painelve property is that expansion \eqref{eq:WTC_expansion} contains three arbitrary functions.

Substituting \eqref{eq:WTC_expansion} into the leading terms of Eq. \eqref{eq:ext_Burgers_3} we get
\begin{equation}
p=-1,\quad u_{0}^{(1,2)}=\frac{-3\pm\sqrt{9-8\bar{\alpha}}}{2\bar{\alpha}} \phi_{x}.
\label{eq:WTC_expansion_1}
\end{equation}
Thus, Eq. \eqref{eq:ext_Burgers_3} admits two expansions in the form \eqref{eq:WTC_expansion}. Substituting the expression
\begin{equation}
u(x,t)=u_{0}^{(1,2)}\,\phi^{-1}+u_{j}\,\phi^{j-1},
\label{eq:WTC_expansion_3}
\end{equation}
into the leading terms of Eq. \eqref{eq:ext_Burgers_3} we find the Fuchs indices
\begin{equation}
\begin{gathered}
j_{1}^{(1)}=-1,\quad j_{2}^{(1)}=3,\quad j_{3}^{(1)}=4+\frac{3\sqrt{9-8\bar{\alpha}}-9}{2\bar{\alpha}},\\
j_{1}^{(2)}=-1,\quad j_{2}^{(2)}=3,\quad j_{3}^{(2)}=4-\frac{3\sqrt{9-8\bar{\alpha}}+9}{2\bar{\alpha}},
\label{eq:WTC_expansion_5}
\end{gathered}
\end{equation}
At this step the necessary condition for Eq. \eqref{eq:ext_Burgers_3} to possess the Painlev\'{e} property is that all Fuchs indices are real integers. It can be shown that this is true only for three values of $\bar{\alpha}$: $\bar{\alpha}_{1}=\frac{9}{8}$, $\bar{\alpha}_{2}=-9$ and $\bar{\alpha}_{3}=1$.

At $\bar{\alpha}_{1}=\frac{9}{8}$ from \eqref{eq:WTC_expansion_5} we obtain
\begin{equation}
\begin{gathered}
j_{1}^{(1)}=-1,\quad j_{2}^{(1)}=3,\quad j_{3}^{(1)}=0,\\
j_{1}^{(2)}=-1,\quad j_{2}^{(2)}=3,\quad j_{3}^{(2)}=0.
\end{gathered}
\end{equation}
In this case Eq. \eqref{eq:ext_Burgers_3} does not possess the Painlev\'{e} property since the coefficient $u_{0}$ is not an arbitrary function.

In the case of $\bar{\alpha}_{2}=-9$ Fuchs indices  \eqref{eq:WTC_expansion_5} have the form
\begin{equation}
\begin{gathered}
j_{1}^{(1)}=-1,\quad j_{2}^{(1)}=3,\quad j_{3}^{(1)}=3,\\
j_{1}^{(2)}=-1,\quad j_{2}^{(2)}=3,\quad j_{3}^{(2)}=6.
\end{gathered}
\end{equation}
We see that there are multiple Fuchs indices. Consequently, expansion \eqref{eq:WTC_expansion} contains logarithmic terms and Eq. \eqref{eq:ext_Burgers_3} does not possess the Painlev\'{e} property at $\bar{\alpha}_{2}=-9$.

At $\bar{\alpha}_{3}=1$ Fuchs indices \eqref{eq:WTC_expansion_5} have the form
\begin{equation}
\begin{gathered}
j_{1}^{(1)}=-1,\quad j_{2}^{(1)}=3,\quad j_{3}^{(1)}=1,\\
j_{1}^{(2)}=-1,\quad j_{2}^{(2)}=3,\quad j_{3}^{(2)}=-2.
\end{gathered}
\end{equation}
Thus, in this case the necessary condition of the Painlev\'{e} test is held. Indeed, at $\bar{\alpha}_{3}=1$ Eq. \eqref{eq:ext_Burgers_3} is the Sharmo--Tasso--Olver equation\cite{Olver1977,Tasso1996}  which is linearizable by the Cole--Hopf transformation.

In this section we have found that Eq. \eqref{eq:ext_Burgers_3} passes the Painleve test only in one case: $\bar{\alpha}=1$. At any other value of $\bar{\alpha}$ the equation does not have the Painleve property. This fact indicates that Eq. \eqref{eq:ext_Burgers_3} is integrable only at $\bar{\alpha}=1$ and this is really the case.

\section{\label{sec:3}Symmetry analysis}

Let us study symmetries and symmetry reductions of Eq. \eqref{eq:ext_Burgers_3}. Below we consider the application of classical and nonclassical methods for finding symmetries of Eq. \eqref{eq:ext_Burgers_3}.

\subsection{Classical Lie method}

First we apply the classical Lie method (see e.g., Refs. \cite{Ovsiannikov1982,Olver1993,Ibragimov2001}) to Eq. \eqref{eq:ext_Burgers_3}. In accordance with this method we consider infinitesimal transformations in (x,t,u) given by
\begin{equation}
\tilde{x}=x+a\,\xi(x,t,u), \quad \tilde{t}=t+a\,\tau(x,t,u),\quad \tilde{u}=u+a\,\eta(x,t,u),
\end{equation}
where $a$ is the group parameter. Eq. \eqref{eq:ext_Burgers_3} is invariant under action of these transformations if the determining equations are satisfied on the solutions of Eq. \eqref{eq:ext_Burgers_3}:
\begin{equation}
X^{(3)}\Delta\Big{|}_{\Delta=0}=0, \quad \Delta(u)=u_{t}-3(u u_{x})_{x}+3\bar{\alpha}\,u^{2}u_{x}+u_{xxx}.
\label{eq:determining_equation}
\end{equation}
Here $X^{(3)}$ is the prolonged infinitesimal generator\cite{Ovsiannikov1982,Olver1993,Ibragimov2001}
\begin{equation}
X^{3}=\xi \partial_{x}+\tau \partial_{t} +\eta \partial_{u}+\eta^{t}\partial_{u_{t}}+\eta^{x}\partial_{u_{x}}+\eta^{xx}\partial_{u_{xx}}+
\eta^{xxx}\partial_{u_{xxx}}.
\end{equation}

Substituting expressions for $\eta^{t},\eta^{x},\eta^{xx},\eta^{xxx}$ into \eqref{eq:determining_equation} we obtain an overdetermined system of linear partial differential equations for the infinitesimals $\xi$, $\tau$, $\eta$. This system of equations has the general solution
\begin{equation}
\xi=c_{1}x+c_{3},\quad \tau=3c_{1}t+c_{2}, \quad \eta=-c_{1}u,
\end{equation}
where $c_{1}$, $c_{2}$, $c_{3}$ are arbitrary constants.

Consequently, Eq. \eqref{eq:ext_Burgers_3} is invariant under three one--parameter Lie groups with the following infinitesimal generators
\begin{equation}
X_{1}=\partial_{x},\quad X_{2}=\partial_{t},\quad
X_{3}=x \partial_{x}+3t\partial_{t}-u\partial_{u}.
\label{eq:ext_Burgers_groups}
\end{equation}

The first infinitesimal generator corresponds to space translation invariance, the second infinitesimal generator corresponds to time translation invariance and the third one corresponds to scaling invariance. Thus there are two classical symmetry reductions of Eq. \eqref{eq:ext_Burgers_3}. The first one is the traveling wave reduction corresponding to $X_{1}+X_{2}$ and the second one is the self--similar reduction corresponding to $X_{3}$. In the next section we will study these symmetry reductions.

Let us note that we cannot indicate the integrable case of Eq. \eqref{eq:ext_Burgers_3} using symmetry analysis presented above. One can find the integrable case of Eq. \eqref{eq:ext_Burgers_3} studying classical symmetries for the potential form of this equation. Indeed, using the variable $u=v_{x}$ in Eq. \eqref{eq:ext_Burgers_3} we find that at $\bar{\alpha}=1$ this equation admits the infinitesimal generator $X_{\infty}=e^{v}\,\psi\partial_{v}$, where $\psi$ is any solution of the third order linear evolution equation $\psi_{t}+\psi_{xxx}=0$.

\subsection{Nonclassical method}

It is known that there are symmetry reductions of partial differential equations which cannot be obtained with the help of the classical Lie method. Symmetries associated with such symmetry reductions are called nonclassical symmetries \cite{Bluman1969,Nucci1994,Popovych1999,Popovych2008}.  The method for finding nonclassical symmetries for partial differential equations was proposed by Bluman and Cole\cite{Bluman1969}. This method  has been applied for finding symmetry reductions and exact solutions of various partial differential equations (see e.g., Refs. \cite{Winternitz1989,Bruzon2003,Bruzon2009,Gordoa2000,Bruzon2012,Nucci2013,Popovych2013}).

According to the method by Bluman and Cole\cite{Bluman1969} we consider the additional auxiliary equation
\begin{equation}
\xi u_{x}+\tau u_{t}-\eta=0.
\label{eq:BC_1}
\end{equation}
Note that Eq. \eqref{eq:BC_1} is the invariant surface condition associated with the vector field $X=\xi \partial_{x}+\tau \partial_{t}+u \partial_{u}$. Then we look for classical symmetries admitted by Eqs. \eqref{eq:ext_Burgers_3} and \eqref{eq:BC_1} simultaneously.

Let us remark that if $X$ is a nonclassical symmetry generator then $\lambda X$ is a nonclassical symmetry generator as well for any function $\lambda(x,t,u)\neq 0$ (see Refs. \cite{Popovych1999,Popovych2008}). Thus, further we have to consider two cases of nonclassical symmetry generators. The first one is the case of $\tau\neq0$, where without loss of generality we can assume that $\tau=1$. The other one is the case of $\tau=0$, where without loss of generality we can assume that $\xi=1$. The case of  $\tau\neq0$ is called the regular case where we have an overdetermined system of equations for infinitesimals. In the singular case $\tau=0$ we obtain a single partial differential equation for infinitesimal $\eta$. As it was shown by Kunzinger and Popovych\cite{Popovych2008} every solution of this equation generates a family of solutions of a considered equation.

Let us study the case of $\tau\neq0$ (we assume that $\tau=1$). Applying the classical Lie method to Eqs. \eqref{eq:ext_Burgers_3} and \eqref{eq:BC_1} and using Eq. \eqref{eq:BC_1} and its differential consequences for eliminating derivatives involving time and using Eq. \eqref{eq:ext_Burgers_3} for eliminating higher order derivative with respect to $x$ we obtain an overdetermined system of nonlinear partial differential equations for $\xi$ and $\eta$.

Solving this system of equations we find
\begin{equation}
\begin{gathered}
\xi=\frac{x+c_{4}}{3(t-c_{5})}, \quad \eta=-\frac{u}{3(t-c_{5})},
\label{eq:BC_5}
\end{gathered}
\end{equation}
where $c_{4}, c_{5}$ are arbitrary constants. It can be shown that these infinitesimals give classical symmetry. Symmetry reduction of Eq. \eqref{eq:ext_Burgers_3} corresponding to \eqref{eq:BC_5} coincides with the reduction generated by a linear combination of infinitesimal generators $X_{1}$, $X_{2}$ and $X_{3}$.

Let us consider the case of $\tau=0$. In this case without loss of generality we assume $\xi=1$. Applying the classical Lie approach and excluding space derivatives using Eq. \eqref{eq:BC_1} and its differential consequences and Eq. \eqref{eq:ext_Burgers_3} we obtain the determining equation
for infinitesimal $\eta$:
\begin{equation}
\begin{gathered}
\eta^{3}\eta_{uuu}+
3\left[2\bar{\alpha} u -u \eta_{uu}+(\eta_{uu}-2)\eta_{u}+\eta_{xuu}\right]\eta^{2}+\vspace{0.1cm}\\+
3\left[\eta_{x}(\eta_{uu}-3)+(\eta_{u}-2u)\eta_{xu}+3\eta_{xxu}\right]\eta+\vspace{0.1cm}\\
+3\bar{\alpha} u^{2} \eta_{x}+3\eta_{x}\eta_{xu}-3 u\eta_{xx}+\eta_{t}+\eta_{xxx}=0.
\label{eq:BC_9}
\end{gathered}
\end{equation}
We cannot find the general solution of Eq. \eqref{eq:BC_9}. Kunzinger and Popovych\cite{Popovych2008} showed that the problem for construction all solutions of Eq. \eqref{eq:BC_9} is completely equivalent to the problem of the construction of all one-parametric solutions families of Eq. \eqref{eq:ext_Burgers_3}. However every particular solution of Eq. \eqref{eq:BC_9} generates nonclassical symmetry reduction of Eq. \eqref{eq:ext_Burgers_3}. Using this fact we can find exact solutions families of Eq. \eqref{eq:ext_Burgers_3} that are not invariant under classical Lie groups admitted by Eq. \eqref{eq:ext_Burgers_3}.

\begin{table}[t]
\caption{Nonclassical infinitesimals admitted by \eqref{eq:ext_Burgers_3}} \label{t: tab1}
\begin{tabular}{ccc}
\textbf{n} & $\bar{\alpha}$ & $\eta$   \\
 \hline
1 & $\bar{\alpha}=\frac{18}{25}$ & $\eta^{1}=\frac{u}{2x+c_{6}}+\frac{5}{(2x+c_{6})^{2}}$ \\ \hline
2 & $\bar{\alpha}\neq0$ & $\eta^{2}=\frac{1}{4}\left(3\pm\sqrt{9-8\bar{\alpha}}\right)u^{2}+\frac{c_{7}\pm 2x}{3t\sqrt{9-8\bar{\alpha}}\mp 9t\pm 2c_{8}}$ \\ \hline
3 & $\bar{\alpha}\neq0$ & $\eta^{3}=\frac{1}{4}\left(3\pm\sqrt{9-8\bar{\alpha}}\right)u^{2}+c_{9}$ \\ \hline
4 & $\bar{\alpha}=0$ & $\eta^{4}=\frac{3}{2}u^{2}+c_{10}x+c_{11}$
 \\ \hline
5 & $\bar{\alpha}=0$ & $\eta^{5}=\frac{2x-c_{12}}{c_{13}-18\,t}$
\end{tabular}
\end{table}

We seek for solutions of \eqref{eq:BC_9} assuming that $\eta$ is a polynomial in $u$. In this way we find infinitesimals which are presented in Table \ref{t: tab1}, where $c_{i}$, $i=6,\ldots,13$ are arbitrary parameters.

Infinitesimal $\eta^{1}$ corresponds to the rational solution of Eq. \eqref{eq:ext_Burgers_3} at $\bar{\alpha}=18/25$. Using infinitesimal $\eta^{2}$ we can obtain exact solution of Eq. \eqref{eq:ext_Burgers_3} which is expressed via the Airy functions. Below we show that this solution is a generalization of the self--similar solution of Eq. \eqref{eq:ext_Burgers_3}. The third infinitesimal $\eta^{3}$ corresponds to rational and solitary wave solutions of Eq. \eqref{eq:ext_Burgers_3}. Using infinitesimals $\eta^{4}$, $\eta^{5}$ we can obtain rational, solitary wave and Airy function solutions of Eq. \eqref{eq:ext_Burgers_3} at $\bar{\alpha}=0$. We believe that infinitesimals $\eta^{1}$, $\eta^{2}$, $\eta^{5}$ correspond to solutions of Eq. \eqref{eq:ext_Burgers_3} that cannot be found with the help of the classical Lie method. Below we construct exact solutions of Eq. \eqref{eq:ext_Burgers_3} corresponding to infinitesimals $\eta^{1}$, $\eta^{2}$, $\eta^{5}$.

Also one can find additional infinitesimal at $\bar{\alpha}=1$:
\begin{equation}
\eta^{6}=u^{2}+v\,u, \quad \bar{\alpha}=1,
\end{equation}
where function $v$ satisfies Eq. \eqref{eq:ext_Burgers_3} at $\bar{\alpha}=1$.  Using this infinitesimal we can construct auto--Backlund transformations for the integrable case of \eqref{eq:ext_Burgers_3}. It is worth noting that for the first time nonclassical symmetries were used for constructing auto--Backlund transformations by Nucci\cite{Nucci1993}.

In this section we have found classical infinitesimal generators admitted by Eq. \eqref{eq:ext_Burgers_3}. Further we use these infinitesimal generators for constructing traveling wave and self--similar solutions of Eq. \eqref{eq:ext_Burgers_3}. We also obtain several nonclassical symmetry generators admitted by Eq. \eqref{eq:ext_Burgers_3}. Below we use some of these generators to construct solutions of Eq. \eqref{eq:ext_Burgers_3} which are not invariant under classical Lie groups.

\section{\label{sec:4}Reductions and exact solutions}
In this section we study symmetry reductions of Eq. \eqref{eq:ext_Burgers_3}. We construct several families of exact solutions of this equation. Both classical and nonclassical symmetry reductions are considered.

\subsection{Traveling wave solutions}
Let us consider reduction of Eq. \eqref{eq:ext_Burgers_3} corresponding to the linear combination of infinitesimal generators $X_{1}$ and $X_{2}$.  Using in Eq. \eqref{eq:ext_Burgers_3} traveling wave variables $u(x,t)=y(z),\, z=x-C_{0}t$ and integrating the result with respect to $z$ we get
\begin{equation}
C_{1}-C_{0}\,y+\bar{\alpha}y^{3}-3yy_{z}+y_{zz}=0,
\label{eq:ext_Burgers_traweling_wave}
\end{equation}
where $C_{1}$ is an integration constant. Introducing the new variable $v=y_{z}$ we obtain the Abel equation of the second kind
\begin{equation}
vv_{y}=3yv-(\bar{\alpha}y^{3}-C_{0}y+C_{1}).
\label{eq:ext_Burgers_traweling_wave_1}
\end{equation}
Eq. \eqref{eq:ext_Burgers_traweling_wave_1} can be transformed to the canonical form with the help of the substitution $\zeta=3/2 y^{2}$:
\begin{equation}
vv_{\zeta}-v=-\frac{2}{9}\bar{\alpha}\zeta+\tilde{C_{0}}-\tilde{C_{1}}\zeta^{-1/2},
\label{eq:ext_Burgers_traweling_wave_3}
\end{equation}
where $\tilde{C_{0}}=C_{0}/3$, $\tilde{C_{1}}=C_{1}/\sqrt{6}$.

List of solvable Abel equation is presented in handbook by Zaitsev and Polyanin\cite{Polyanin}. General solution of Eq. \eqref{eq:ext_Burgers_traweling_wave_3} at $\tilde{C}_{0},\tilde{C}_{1}\neq0$ can be found only in the case of $\bar{\alpha}=1$.

\begin{figure}[!htb]
\center
 \includegraphics[height=7cm]{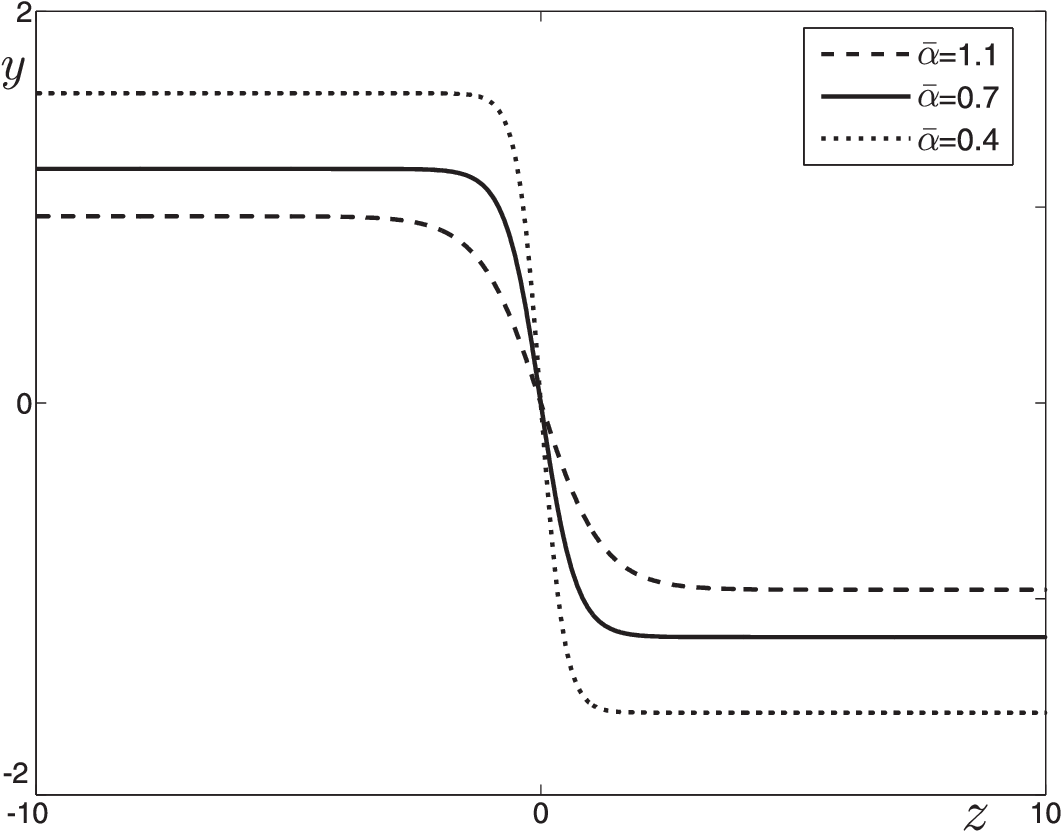}
    \caption{Exact solution \eqref{eq:ext_Burgers_kinks_solution} for various values of $\bar{\alpha}.$}
  \label{fig1}
\end{figure}

For constructing particular solutions of \eqref{eq:ext_Burgers_traweling_wave} we use the simplest equation method\cite{Kudryashov2005}. We find that Eq. \eqref{eq:ext_Burgers_traweling_wave} at $C_{1}=0$ has the following solution
\begin{equation}
y=\frac{-3\pm \sqrt{9-8\bar{\alpha}}}{2\bar{\alpha}}\sqrt{B}\tanh\{\sqrt{B}(z-z_{0})\},\quad
B=\frac{2 C_{0}\bar{\alpha}}{9-3\sqrt{9\mp8\bar{\alpha}}-4\bar{\alpha}}.
\label{eq:ext_Burgers_kinks_solution}
\end{equation}
One can see that solution \eqref{eq:ext_Burgers_kinks_solution} cannot be used at $\bar{\alpha}=0$. In this case Eq. \eqref{eq:ext_Burgers_traweling_wave} admits stationary solution ($C_{0}=0$) at $C_{1}=0$
\begin{equation}
y=-\frac{2}{3}\sqrt{B}\tanh\{\sqrt{B}(z-z_{0})\},
\label{eq:ext_Burgers_kinks_solution_alpha=0}
\end{equation}
where $B$ is an arbitrary constant. Let us note that solutions \eqref{eq:ext_Burgers_kinks_solution}, \eqref{eq:ext_Burgers_kinks_solution_alpha=0} can be obtained by direct integration of Eq. \eqref{eq:ext_Burgers_traweling_wave}.

The plots of solution \eqref{eq:ext_Burgers_kinks_solution} for several values of $\bar{\alpha}$ are presented in Fig.\ref{fig1}. We can see that solution \eqref{eq:ext_Burgers_kinks_solution} is a weak shock wave.

\subsection{Self--similar solutions}

Let us consider the self--similar reduction of Eq. \eqref{eq:ext_Burgers_3}. Using the variables
\begin{equation}
u=C_{2}t^{-1/3}f(\theta),\quad \theta=C_{3}x\,t^{-1/3},
\label{eq:self-similar_variables}
\end{equation}
from Eq. \eqref{eq:ext_Burgers_3} at $C_{2}=-C_{3}=-(3)^{-1/3}$ we get
\begin{equation}
(\theta f)_{\theta}=3(f f_{\theta})_{\theta}+3\bar{\alpha}f^{2}f_{\theta}+f_{\theta\theta\theta}.
\label{eq:ext_Burgers_self_similar_1}
\end{equation}
Integrating \eqref{eq:ext_Burgers_self_similar_1} once we have
\begin{equation}
\theta f=3ff_{\theta}+\bar{\alpha}f^{3}+f_{\theta\theta}+C_{4},
\label{eq:ext_Burgers_self_similar_1}
\end{equation}
where $C_{4}$ is an integration constant. It is known\cite{Polyanin} that the general solution of Eq. \eqref{eq:ext_Burgers_self_similar_1} can be only obtained in the case of $\bar{\alpha}=1$. However at the arbitrary value of $\bar{\alpha}$ Eq. \eqref{eq:ext_Burgers_self_similar_1} admits particular solution which is expressed via the Airy functions.

Indeed, if the following relation holds
\begin{equation}
C_{4}=-\frac{\sqrt{9-8\bar{\alpha}}\pm3}{3\sqrt{9-8\bar{\alpha}}\mp 4\bar{\alpha} \pm 9},
\end{equation}
Eq. \eqref{eq:ext_Burgers_self_similar_1} has the solution in the form
\begin{equation}
\begin{gathered}
f=-\frac{3\pm\sqrt{9-8\bar{\alpha}}}{2\bar{\alpha}} \times \hfill \\ \times \frac{(-B)^{1/3}(C_{5}\mbox{Ai}^{'}\{-(-B)^{1/3}(\theta-\theta_{0})\}+
\mbox{Bi}^{'}\{-(-B)^{1/3}(\theta)\})}{C_{5}\mbox{Ai}\{-(-B)^{1/3}(\theta-\theta_{0})\}+
\mbox{Bi}\{-(-B)^{1/3}(\theta)\}}, \quad \bar{\alpha}\neq 0
\label{eq:ext_Burgers_self_similar_3}
\end{gathered}
\end{equation}
where
\begin{equation}
B=\frac{2\bar{\alpha}}{9\pm3\sqrt{9-8\bar{\alpha}}-4\bar{\alpha}}
\end{equation}
and $'$ denotes the derivative with respect to the function argument, $C_{5}$ is an arbitrary constant.

\begin{figure}[!htb]
\center
 \includegraphics[height=7cm]{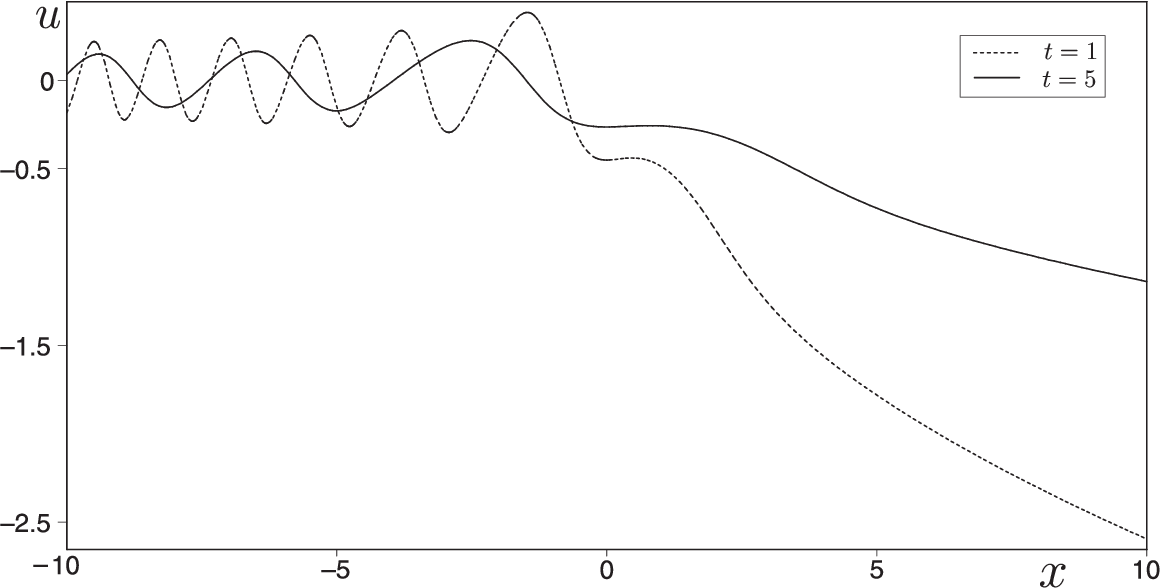}
    \caption{Self--similar solution \eqref{eq:ext_Burgers_self_similar_5} of Eq. \eqref{eq:ext_Burgers_3} for $C_{6}=0.1$ and $C_{7}=1$.}
  \label{fig1a}
\end{figure}

In the case of $\bar{\alpha}=0$ we cannot find self--similar solution of Eq. \eqref{eq:ext_Burgers_3}. However, one can obtain stationary solution of Eq. \eqref{eq:ext_Burgers_3} at $\bar{\alpha}=0$ that is expressed via the Airy functions. For instance, this solution can be obtained using nonclassical infinitesimal $\eta^{4}$.

Let us note that using the Cole--Hopf transformation one can obtain more general self--similar solution of Eq. \eqref{eq:ext_Burgers_3} at $\bar{\alpha}=1$. This solution has the form
\begin{equation}
\begin{gathered}
f=\frac{\Psi_{\theta}}{\Psi},\quad \Psi=\int (C_{6} \mbox{Ai}(\theta)+\mbox{Bi}(\theta))d\theta+C_{7},
\label{eq:ext_Burgers_self_similar_5}
\end{gathered}
\end{equation}
where $C_{6}$ and $C_{7}$ are arbitrary constants.

Using \eqref{eq:self-similar_variables} and \eqref{eq:ext_Burgers_self_similar_5} we can find dependence of  \eqref{eq:ext_Burgers_self_similar_5} on $x$ and $t$. The plot of this solution at various values of $t$ is presented in Fig.\ref{fig1a}. From Fig.\ref{fig1a} we see that this solution is a weak shock wave with oscillations on a wave crest. When time $t$ increases the number of oscillations and the amplitude of waves decrease and the width of wave front increases. Waves of this type were observed experimentally in a liquid with gas bubbles.

\subsection{Nonclassical reductions and their exact solutions}

Let us study nonclassical symmetry reductions of Eq. \eqref{eq:ext_Burgers_3}. First we consider reduction corresponding to nonclassical infinitesimal $\eta^{1}$ from Table \ref{t: tab1}.

Using $\eta^{1}$ we find similarity variables for Eq. \eqref{eq:ext_Burgers_3} at $\bar{\alpha}=\frac{18}{25}$
\begin{equation}
u(x,t)=\sqrt{2x+c_{6}}\,h(t)-\frac{5}{3(2x+c_{6})},
\label{eq:nc_reduction_1}
\end{equation}
where function h(t) satisfies the equation
\begin{equation}
h_{t}+\frac{54}{25}h^{3}=0
\end{equation}
Solving this equation and using \eqref{eq:nc_reduction_1} we obtain a rational solution of Eq. \eqref{eq:ext_Burgers_3} at $\bar{\alpha}=\frac{18}{25}$
\begin{equation}
u(x,t)=\frac{5}{3}\left(\sqrt{\frac{2x+c_{6}}{12\,t+t_{0}}}-\frac{1}{2x+c_{6}}\right),
\label{eq:nc_reduction_3}
\end{equation}
where $t_{0}$ is an arbitrary constant. We see that we have obtained an exact solution of Eq. \eqref{eq:ext_Burgers_3} that cannot be found using the classical Lie method.

Let us consider the nonclassical infinitesimal $\eta^{5}$ from Table \ref{t: tab1}. In this case symmetry reduction has the form
\begin{equation}
u(x,t)=\frac{x^{2}-c_{12}x}{2c_{13}-18t}+h(t),
\label{eq:nc_reduction_5}
\end{equation}
where $h(t)$ satisfies the equation
\begin{equation}
h_{t}=\frac{1}{4(c_{13}-9t)^{2}}\left(12(c_{13}-9\,t)h+3c_{12}^{2}\right).
\label{eq:nc_reduction_7}
\end{equation}
Using \eqref{eq:nc_reduction_5}, \eqref{eq:nc_reduction_7} we find the following solution of Eq. \eqref{eq:ext_Burgers_3} at $\bar{\alpha}=0$
\begin{equation}
u=\frac{(2x-c_{12})^{2}}{8(c_{13}-9t)}+\frac{C_{8}}{(c_{13}-9t)^{1/3}},
\end{equation}
where $C_{8}$ is an integration constant.

Nonclassical infinitesimals $\eta^{3}$ and $\eta^{4}$ correspond to the traveling wave and stationary solutions that have been obtained above. It is interesting to consider nonclassical infinitesimal $\eta^{2}$, nonclassical infinitesimal $\eta^{3}$ at $c_{9}=0$ and nonclassical infinitesimal $\eta^{4}$ at $c_{10}=c_{11}=0$.

Nonclassical infinitesimal $\eta^{2}$ corresponds to the exact solution of Eq. \eqref{eq:ext_Burgers_3} that is expressed via the Airy functions and has the form
\begin{equation}
\begin{gathered}
u=\left(\frac{3t(-3+\sqrt{9-8\bar{\alpha}})+2c_{8}}{[c_{8}(3+\sqrt{9-8\bar{\alpha}})-12\bar{\alpha} t]^{2}}\right)^{1/3}\frac{C_{9}\mbox{Ai}^{'}(\theta)+\mbox{Bi}^{'}(\theta)}
{C_{9}\mbox{Ai}(\theta)+\mbox{Bi}(\theta)},\vspace{0.1cm}\\
\theta=\left(x+c_{7}\right)\left(\frac{3t(-3+\sqrt{9-8\bar{\alpha}})+2c_{8}}
{\sqrt{c_{8}(3+\sqrt{9-8\bar{\alpha}})-12\bar{\alpha} t}}\right)^{-\frac{2}{3}},
\label{eq:nc_reduction_9}
\end{gathered}
\end{equation}
where $'$ denotes the derivative with respect to function argument, $C_{9}$ is an arbitrary constant. Solution \eqref{eq:nc_reduction_9} is a generalization of self--similar solution \eqref{eq:ext_Burgers_self_similar_3} and coincides with this solution at $c_{7}=c_{8}=0$.

Using nonclassical infinitesimal $\eta^{3}$ at $c_{9}=0$ and nonclassical infinitesimal $\eta^{4}$ at $c_{10}=c_{11}=0$ we can obtain the following rational solutions of Eq. \eqref{eq:ext_Burgers_3}
\begin{equation}
\begin{gathered}
u=-\frac{4}{(3\pm \sqrt{9-8\bar{\alpha}})(x-x_{0})}, \quad \bar{\alpha}\neq0,\vspace{0.1cm}\\
u=-\frac{2}{3(x-x_{0})}, \quad \bar{\alpha}=0,
\end{gathered}
\end{equation}
where $x_{0}$ is an arbitrary constant. Thus, we have found one--parametric families of stationary rational solutions of Eq. \eqref{eq:ext_Burgers_3}.

In this section we have found several families of exact solutions of Eq. \eqref{eq:ext_Burgers_3}. Two families of exact solutions correspond to the classical symmetry reductions of Eq. \eqref{eq:ext_Burgers_3}. The other families of solutions correspond to the nonclassical symmetry reductions of Eq. \eqref{eq:ext_Burgers_3}. We believe that all found exact solutions are new.

\section{\label{sec:5}Numerical investigation}
As it has been shown above Eq. \eqref{eq:ext_Burgers_3} is integrable only at one value of the bifurcation parameter ($\bar{\alpha}=1$). Consequently, at this value of $\bar{\alpha}$ one can obtain a plenty of exact solutions. There is an interesting question. What happens with the solutions of the integrable case of Eq. \eqref{eq:ext_Burgers_3} when the value of $\bar{\alpha}$ is being varied?
In other words, it is interesting to study the stability of solutions for Eq. \eqref{eq:ext_Burgers_3} with respect to perturbations of the parameter $\bar{\alpha}$. For this purpose we will use the numerical approach.

Let us consider the boundary value problem for Eq. \eqref{eq:ext_Burgers_3} with periodic boundary conditions and some initial condition. We use the integrating factor with the
fourth--order Runge--Kutta approximation method for the numerical calculations (see Refs. \cite{Cox2002,Trefethen2005,Kudryashov2011}).

We can present the perturbed Burgers equation in the form
\begin{equation}
u_{t}=L(u)+N(u),
\label{eq:numerical_1}
\end{equation}
where
$L(u)=-u_{xxx}$ and  $N(u)=3(u u_{x})_{x}-3\bar{\alpha}u^{2}u_{x}$. The boundary conditions and the initial condition have the form
\begin{equation}
\begin{gathered}
u_{ix}(x,t)=u_{ix}(x+H,t), \quad i=0,1,2 \hfill \\
u(x,0)=u_{0}(x), \hfill
\label{eq:numerical_3}
\end{gathered}
\end{equation}
where $u_{ix}=\partial^{i} u/ \partial x^{i}$ and $H$ is a period.

We discretize the spatial part of \eqref{eq:numerical_1} using the Fourier transform. As a result we obtain the system of ordinary differential equations
\begin{equation}
\hat{u}_{t}=\hat{L}(\hat{u})+\hat{N}(\hat{u}), \quad \hat{u}\left.|\right._{t=0}=\hat{u}_{0},
\label{eq:numerical_5}
\end{equation}
where $\hat{u}$, $\hat{L}$ and $\hat{N}$ are the Fourier transforms of $u$, $L$ and $N$ respectively.

Applying the integrating factor in \eqref{eq:numerical_5}
\begin{equation}
\hat{u}=e^{\hat{L}\,t}v,
\end{equation}
we have
\begin{equation}
v_{t}=e^{-\hat{L}\,t}N(e^{\hat{L}\,t}v).
\label{eq:numerical_7}
\end{equation}
We solve this system of ordinary differential equations using the fourth-order Runge--Kutta
method.

\begin{figure}[!ht]
\begin{center}
 \includegraphics[height=6cm]{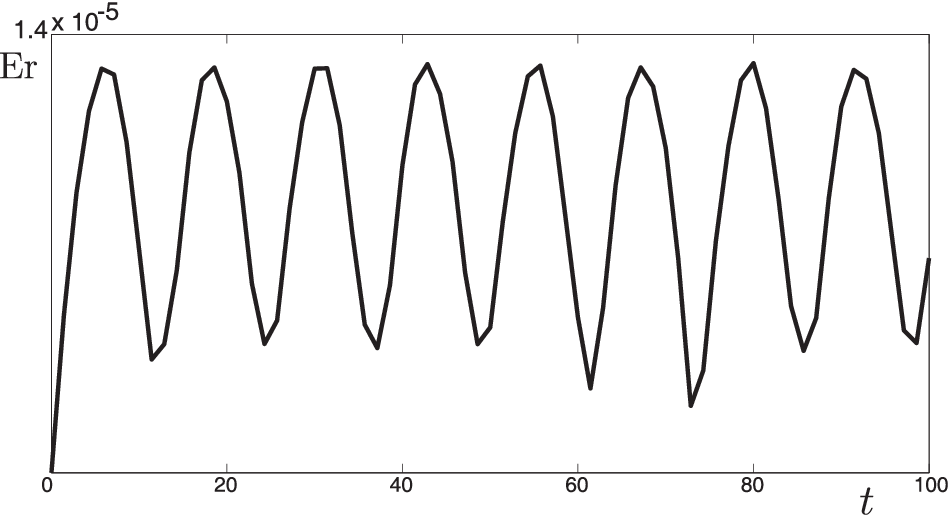}
    \caption{The dependence of average error \eqref{eq:numerical_11} on time for $N=256$.}
      \label{fff}
\end{center}
\end{figure}

To verify our numerical algorithm we use one of the periodic exact solutions of Eq. \eqref{eq:ext_Burgers_3} at $\bar{\alpha}=1$. This solution has the form
\begin{equation}
u=\frac{k\sin(k\,x+k^{3}\,t-\phi_{0})}{C_{10}+\cos(k\,x+k^{3}\,t-\phi_{0})},
\label{eq:numerical_9}
\end{equation}
where $k$, $C_{10}$ and $\phi_{0}$ are arbitrary constants.

We calculate the average error of our numerical algorithm by the formula
\begin{equation}
\mbox{Er}=\frac{1}{N}\sum\limits_{i=1}^{N}|u^{i}_{\mbox{exact}}-u^{i}_{\mbox{num}}|,
\label{eq:numerical_11}
\end{equation}
where $N$ is the number of approximation points on the $x$ axis.

\begin{figure}[!ht]
\begin{center}
 \includegraphics[width=12cm]{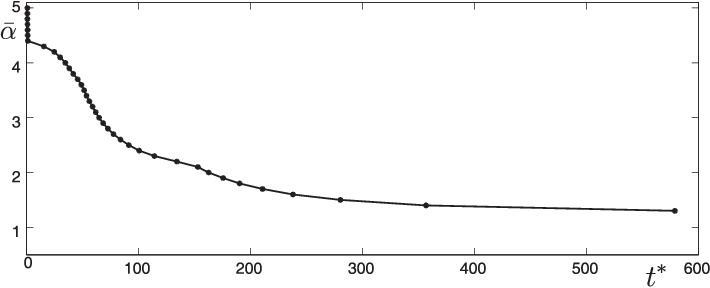}
    %\caption{The dependence of $\bar{\alpha}$ from time $t^{*}$.}
    \caption{The values of $\bar{\alpha}$ versus $t^{*}$}
      \label{fig3}
\end{center}
\end{figure}

We use solution \eqref{eq:numerical_9} at $t=0$ as the initial condition for problem \eqref{eq:numerical_3}. We calculated the average error by formula \eqref{eq:numerical_11} for $N=256$ and $N=512$. Total time of calculations was chosen equal to $10$. We obtain that the magnitude of the average error in both cases is very small and equals to $10^{-10}$ and to $10^{-14}$ respectively. Thus we can use $N=256$ in all other calculations. We have also tested our numerical algorithm for the long time of calculations. The dependence of average error \eqref{eq:numerical_11} on time $t$ is presented in Fig.\ref{fff}.

Let us investigate the stability of solution \eqref{eq:numerical_9} under the perturbations of parameter $\bar{\alpha}$. We again use this solution as the initial condition and perform calculations for $\bar{\alpha}>1$. The calculations are performed until numerical solution does not break up. In this way for each value of $\bar{\alpha}$ we can obtain the value of time $t=t^{*}$ at which the solution breaks up. The values of $\bar{\alpha}$ versus $t^{*}$ is presented in Fig. \ref{fig3}. From Fig.\ref{fig3} we see that there is exist a critical value of parameter $\bar{\alpha}$ at which time $t^{*}$ becomes very small. This critical value $\bar{\alpha}^{*}$ is located between $4.3$ and $4.4$.

We performed analogous calculations for $\bar{\alpha}\in [0,1)$. We have found that solution remains stable until $t\sim 500$ for all $\bar{\alpha}\in [0,1)$, thus there is no critical value of $\bar{\alpha}$ on this interval.

\begin{figure}[!ht]
\begin{center}
 \includegraphics[width=14cm]{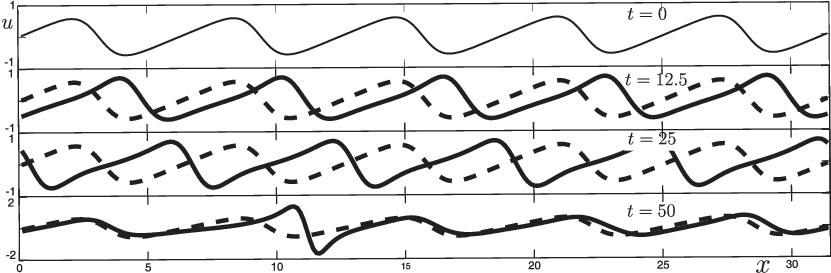}
    \caption{Evolution of solution \eqref{eq:numerical_9} governed by Eq. \eqref{eq:ext_Burgers_3} for $\bar{\alpha}=3.57$.}
      \label{fig4}
\end{center}
\end{figure}

In Fig.\ref{fig4} we demonstrate evolution of solution \eqref{eq:numerical_9} governed by Eq. \eqref{eq:ext_Burgers_3} at $\bar{\alpha}=3.57$. We see that the solution remains close to the exact solution until time $t=12.5$. On the next stage of evolution the numerical solution is slightly deformed in comparison to the exact solution. At $t=50$ the numerical solution significantly changes the form. However, it remains stable.

From Figs.\ref{fig3},\ref{fig4} we see that periodic exact solution \eqref{eq:numerical_9} is stable under the variations of parameter $\bar{\alpha}$. Thus under small variations of $\bar{\alpha}$ we can use solution  \eqref{eq:numerical_9} for the description of waves governed by the nonintegrable case of Eq. \eqref{eq:ext_Burgers_3}.

\begin{figure}[!ht]
\begin{center}
 \includegraphics[width=14cm]{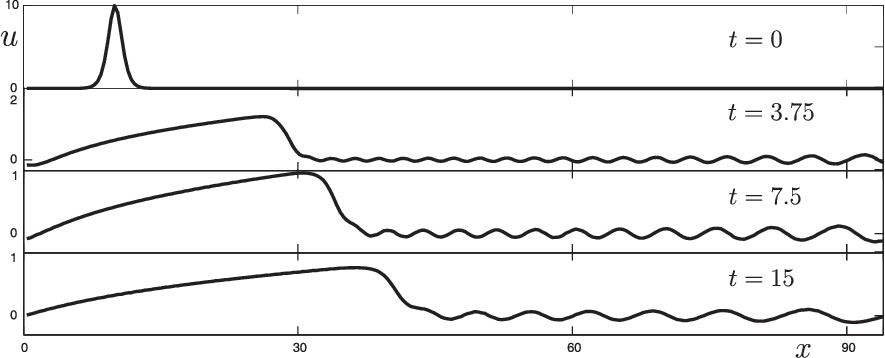} %height=6cm
    \caption{Evolution of a solitary wave in water with carbon dioxide bubbles governed by Eq. \eqref{eq:ext_Burgers_3} at $\bar{\alpha}=1.12$.}
      \label{f5}
\end{center}
\end{figure}

Let us study the propagation of solitary waves governed by Eq. \eqref{eq:ext_Burgers_3} at values of the parameter $\bar{\alpha}$ typical for a real liquid with gas bubbles. We consider waves propagated in water with carbon dioxide bubbles with equilibrium bubbles radius equal to $R_{0}=2.8\,10^{-5}$m. In this case the parameter $\bar{\alpha}$ has the value $\bar{\alpha}\simeq 1.12$.

\begin{figure}[!htp]
\begin{center}
 \includegraphics[width=14cm]{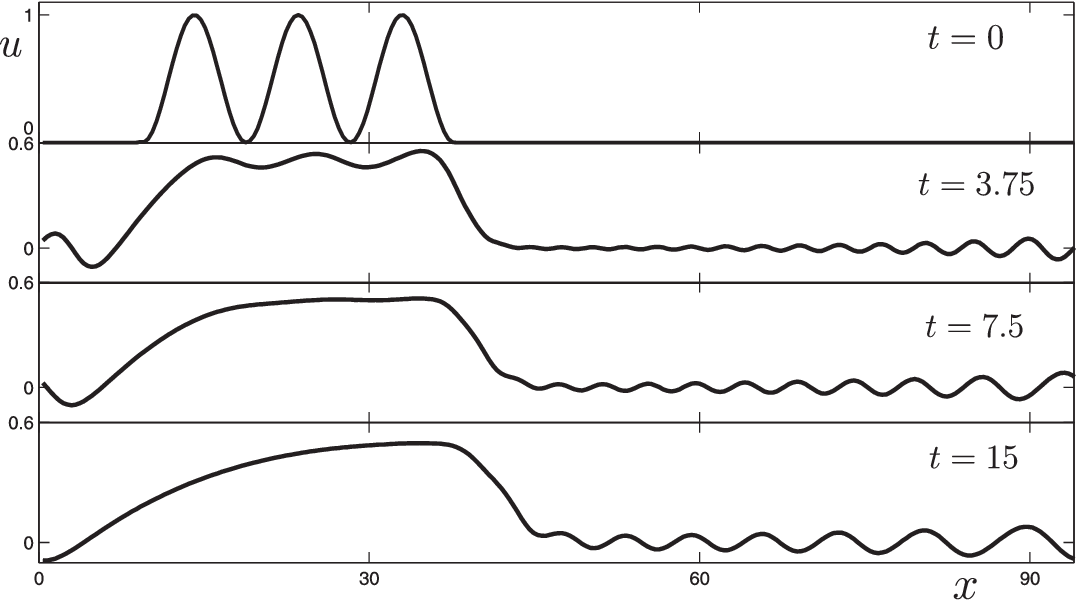} %height=6cm,
    \caption{Evolution of a compact wave in water with carbon dioxide bubbles governed by \eqref{eq:ext_Burgers_3} at $\bar{\alpha}=1.12$.}
      \label{f6}
\end{center}
\end{figure}

In Fig.\ref{f5} we demonstrate the evolution of initial solitary wave $u_{0}=10\cosh^{-2}\{x-10\}$ governed by Eq. \eqref{eq:ext_Burgers_3}. We see that the solitary wave is transformed to the weak shock wave with the oscillating tail after the wave crest.

Let us consider as the initial condition the wave with compact support:
\begin{equation}
u_{0}= \left\{
\begin{gathered}
\sin^{2}(x/3), \, \quad x\in [3\pi,12\pi]  \hfill \\
0, \quad \quad \quad \quad\,\,\,\, x \not\in [3\pi,12\pi] \hfill
\end{gathered}
\right.
\label{eq:alpha}
\end{equation}
From Fig.\ref{f6} we see that this wave is transformed to a weak shock waves with the oscillating tail after the wave crest as well. We perform analogous calculations for initial Gaussian and rectangular pulses. We obtained that these initial pulses evolve to the same weak shock wave.

\begin{figure}[!ht]
\begin{center}
 \includegraphics[width=10cm]{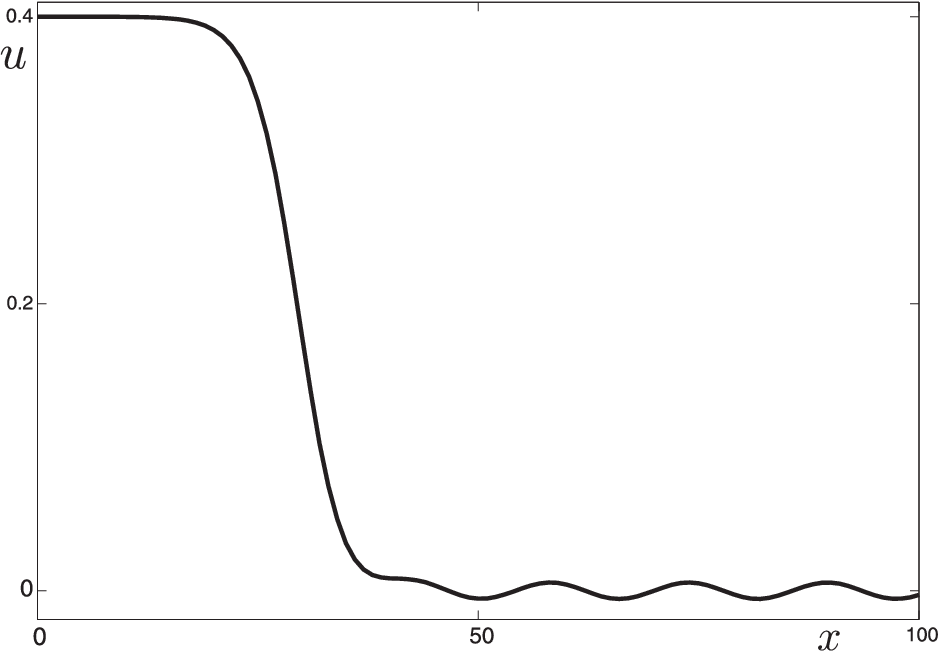} 
    \caption{Exact solution \eqref{eq:numerical_15} of the integrable case of Eq. \eqref{eq:ext_Burgers_3} at $k=0.4$, $C_{11}=70$, $t=15$, $\phi_{0}^{(1)}=\phi_{0}^{(2)}=15$. }
      \label{f7}
\end{center}
\end{figure}

We can see that several types of solitary wave initial conditions evolve to the same asymptotic profile. We found that solutions of Eq. \eqref{eq:ext_Burgers_3} have the same asymptotic behavior at large values of $t$. Thus there is a stable solution of Eq. \eqref{eq:ext_Burgers_3} that describes dynamics of solitary waves at large values of $t$.

Let us note that the integrable case of Eq. \eqref{eq:ext_Burgers_3} admits analytical solution that is very similar to this asymptotic profile. This solution has the form
\begin{equation}
u=-\frac{\Psi_{x}}{\Psi}, \quad \Psi=C_{11}+\cos(kx+k^{3}t+\phi_{0}^{(1)})+\exp(-kx+k^{3}t+\phi_{0}^{(2)}),
\label{eq:numerical_15}
\end{equation}
where $k,C_{11},\phi_{0}^{(1)}$ and $\phi_{0}^{(2)}$ are arbitrary constants. The plot of solution \eqref{eq:numerical_15} at $t=15$ is presented in Fig.\ref{f7}. From Fig.\ref{f7} we see that solution \eqref{eq:numerical_15} is very similar to numerical solutions obtained above.

\section{\label{sec:6} Final remarks and conclusions}

The perturbed Korteweg--de Vries equation has been investigated. We have shown that this equation passes the Painleve test only in the case of $\bar{\alpha}=1$. This case correspond to the integrable Sharma--Tasso--Olver equation. We have applied the classical Lie method to the perturbed Korteweg--de Vries equation. Three classical infinitesimal generators admitted by this equation have been found: translations in $x$ and $t$ and scaling in $x$, $t$ and $u$. Some nonclassical symmetries of the perturbed Korteweg--de Vries equation have been also obtained using the method by Bluman and Cole. Exact solutions of classical and nonclassical symmetry reductions have been constructed. We have obtained traveling wave solutions, self--similar solutions and three families of solutions that are invariant under nonclassical symmetries. We believe that all found exact solutions are new.

The stability of exact solutions of the perturbed Korteweg--de Vries equation was investigated numerically. We have shown that periodic solution \eqref{eq:numerical_11} is stable under the variation of parameter $\bar{\alpha}$. We have seen that dynamics of waves described by nonintegrable case of Eq. \eqref{eq:ext_Burgers_3} remains close to the dynamics of waves described by integrable case of Eq. \eqref{eq:ext_Burgers_3} at small variations of parameter $\bar{\alpha}$. It can be conjectured that other periodic solutions of the integrable case of Eq. \eqref{eq:ext_Burgers_3} are also stable under variations of $\bar{\alpha}$.

We have found that the dynamics of waves governed by Eq. \eqref{eq:ext_Burgers_3} at large values of time is the same for various initial conditions. The initial disturbances evolve to the same weak shock wave with oscillations after wave crest.  Consequently, for large values of time there is a stable solution of Eq. \eqref{eq:ext_Burgers_3}. We have shown that integrable case of Eq. \eqref{eq:ext_Burgers_3} admits an analytical solution in the from of a weak shock wave with oscillations after wave crest.

This research was partially supported by grant for Scientific Schools 2296.2014.1., by grant for the state support of young Russian scientists 3694.2014.1 and by grant of Russian Science Foundation 14--11--00258.

\end{document}